\documentclass[twocolumn,english,pra,aps,superscriptaddress,longbibliography]{revtex4-1}
\usepackage[T1]{fontenc}
\usepackage{mathrsfs}
\usepackage{amsmath}
\usepackage{graphicx}

\makeatletter
\usepackage{babel}

\makeatother

\usepackage{babel}
\begin{document}
\title{Boosting the performance of the Quantum Otto heat engines}
\date{\today}
\author{Jin-Fu Chen}
\address{Beijing Computational Science Research Center, Beijing 100193, China}
\address{Graduate School of China Academy of Engineering Physics, No. 10 Xibeiwang
East Road, Haidian District, Beijing, 100193, China}
\author{Chang-Pu Sun}
\address{Beijing Computational Science Research Center, Beijing 100193, China}
\address{Graduate School of China Academy of Engineering Physics, No. 10 Xibeiwang
East Road, Haidian District, Beijing, 100193, China}
\author{Hui Dong}
\email{hdong@gscaep.ac.cn}

\address{Graduate School of China Academy of Engineering Physics, No. 10 Xibeiwang
East Road, Haidian District, Beijing, 100193, China}
\begin{abstract}
To optimize the performance of a heat engine in finite-time cycle,
it is important to understand the finite-time effect of thermodynamic
processes. Previously, we have shown that extra work is needed to
complete a quantum adiabatic process in finite time, and proved that
the extra work follows a $\mathcal{C}/\tau^{2}$ scaling for long
control time $\tau$. There the oscillating part of the extra work
is neglected due to the complex energy-level structure of the particular
quantum system. However, such oscillation of the extra work can not
be neglected in some quantum systems with simple energy-level structure,
e. g. the two-level system or the quantum harmonic oscillator. In
this paper, we build the finite-time quantum Otto engine on these
simple systems, and find that the oscillating extra work leads to
a jagged edge in the constraint relation between the output power
and the efficiency. By optimizing the control time of the quantum
adiabatic processes, the oscillation in the extra work is utilized
to enhance the maximum power and the efficiency. We further design
special control schemes with the zero extra work at the specific control
time. Compared to the linear control scheme, these special control
schemes of the finite-time adiabatic process improve the maximum power
and the efficiency of the finite-time Otto engine.
\end{abstract}
\maketitle

\section{Introduction}

Quantum thermodynamics \citep{Maruyama2009,Esposito_2009,Campisi_2011,Strasberg_2017,Vinjanampathy_2016}
studies the effect of quantum characteristics, e. g. coherence \citep{Scully_2003,Quan2006,Solinas2016,Francica2019},
entanglement \citep{Benatti2003,Kraus2008,Altintas2014,Tavakoli2019},
and quantum many-body effect \citep{Jaramillo2016,Ma2017PhysRevE96_22143,Bengtsson2018,Chen2018}
on the thermodynamic property of the system. One important topic is
to find quantum heat engines as counterparts of the classical ones.
To design a practical heat engine with non-zero output power, the
finite-time quantum thermodynamics \citep{Curzon_1975,Salamon1983,Esposito_2010,Andresen2011,Whitney2014PhysRevLett112_130601,Dann2019}
needs to be studied instead of quasi-static thermodynamics \citep{Quan_2007,Ma2017PhysRevE96_22143,Bengtsson2018,Chen2018}.
Therefore understanding the finite-time effect of thermodynamic processes
is crucial to the optimization of the finite-time heat engine \citep{Tu2008JPhysAMathTheor41_312003,Rutten2009,Whitney2014PhysRevLett112_130601,Alecce2015,Shiraishi2016PhysRevLett117_190601,Deffner2018}.
Based on the universal $\mathcal{C}/\tau$ scaling of the entropy
production in finite-time isothermal processes \citep{Salamon1983},
the efficiency at the maximum power is obtained analytically for the
finite-time Carnot-like engine \citep{Curzon_1975,Schmiedl2007,Esposito_2010,Cavina2017}.
The trade-off relation between efficiency and power is further established
recently \citep{Shiraishi2016PhysRevLett117_190601,Ryabov2016PhysRevE93_50101,Long2016,Holubec2016,Ma2018,Ma2018PhysRevE98_42112}
for finite-time Carnot cycle. The finite-time heat engine of other
types, e. g. the finite-time Otto engine, has been studied \citep{Abah2012,Rosnagel2014,Alecce2015,Karimi_2016,Insinga2016,Campisi2016NatCommun7_,Kosloff2017,Deffner2018,1812.05089,Denzler2019}
and is shown with better performance by the technique of the shortcut
to adiabatic \citep{Chen2010PhysRevLett104_63002,Deng2013,Abah2018,Deng2018SciAdv4_5909,Cakmak2019}.
Yet, the optimization of the finite-time Otto engine lacks a general
principle compared to the universal $\mathcal{C}/\tau$ scaling of
the entropy production in the finite-time Carnot-like engine.

Evaluating the finite-time effect of the adiabatic processes is the
key to the optimization of the finite-time Otto engine, which consists
two adiabatic processes and two isochoric processes. We consider the
situation where the time consuming of the finite-time isochoric processes
can be neglected compared to the finite-time adiabatic processes \citep{Chotorlishvili2016,Abah2018}.
During the finite-time adiabatic process, the system is isolated from
the environment and evolves under the time-dependent Hamiltonian \citep{Su2018}.
When energy levels of different states do not cross, the quantum adiabatic
approximation is valid for long control time \citep{Quan_2007}. In
this situation, the theorem of high-order adiabatic approximation
provides a perturbative technique to derive the finite-time correction
to higher orders of the inverse control time \citep{Sun_1988,Wilczek1989_,Sun1990,Rigolin_2008}.
It requires positive extra work to complete the adiabatic process
in finite time.

In our previous paper \citep{Chen2019}, we find that the extra work
in the finite-time adiabatic process can be naturally divided into
the mean extra work and the oscillating extra work. With the increasing
control time $\tau$, the mean extra work decreases monotonously,
obeying a general $\mathcal{C}/\tau^{2}$ scaling behavior. The oscillating
extra work oscillates around zero for larger $\tau$, and is neglected
due to the incommensurable energy of different states in large systems.
Yet, this oscillating extra work can not be neglected for the system
with simple energy-level structure. In this paper, we continue the
study of the oscillating extra work, and show its effects on some
simple systems, such as the two-level system and the quantum harmonic
oscillator. We find that the oscillation of the extra work can be
utilized to enhance the output power of the heat engine. Besides,
we obtain special control schemes of the adiabatic processes with
zero extra work at the specific control time. The special control
scheme further improves the maximum power of the Otto engine.

This paper is organized as follows. In Sec. \ref{sec:Finite-time-Quantum-Otto},
we review the generic finite-time quantum Otto engine, and list the
dependence of the power and the efficiency on the extra work in the
finite-time adiabatic processes for later discussion. In Sec. \ref{sec:Two-level-Otto-heat}
and \ref{sec:Quantum-Harmonic-Otto}, the finite-time quantum Otto
cycles on two-level system and quantum harmonic oscillator are studied,
respectively. The conclusion is given in Sec. \ref{sec:Conclusion}.

\section{Finite-time Quantum Otto engine\label{sec:Finite-time-Quantum-Otto}}

\begin{figure}
\includegraphics[width=4cm]{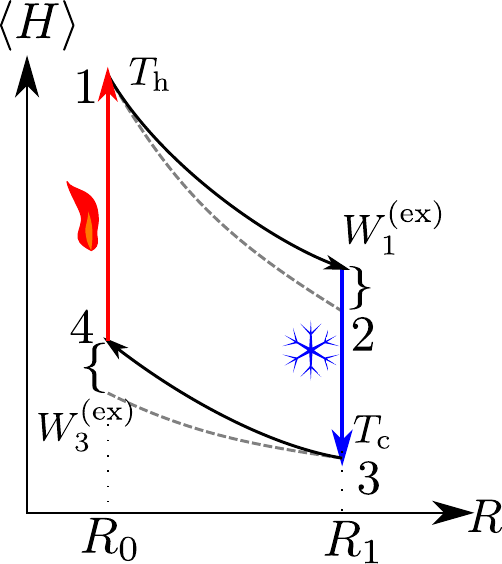}

\caption{Energy-parameter ($\left\langle H\right\rangle -R$) diagram of the
finite-time Otto cycle. The solid line with arrows presents the finite-time
cycle, where two vertical colored lines present the isochoric processes
(red for heating and blue for cooling), and two black lines present
the finite-time adiabatic processes. The dashed lines present the
quasi-static adiabatic processes. The extra work in the two finite-time
adiabatic processes is marked as $W_{1}^{(\mathrm{ex})}(\tau_{1})$
and $W_{3}^{(\mathrm{ex})}(\tau_{3})$ . \label{fig:Finite-time-Otto-cycle.}}
\end{figure}

In this section, we briefly review the finite-time Otto cycle. A generic
finite-time Otto cycle consists four strokes, two finite-time adiabatic
processes and two finite-time isochoric processes, illustrated on
$\left\langle H\right\rangle -R$ diagram in Fig. \ref{fig:Finite-time-Otto-cycle.}.

In the finite-time adiabatic processes ($1\rightarrow2$ and $3\rightarrow4$)
with the control time $\tau_{1}$ and $\tau_{3}$, the work is performed
by tuning the parameter $R(t)$ in the Hamiltonian $H(t)\equiv H[R(t)]$,
from $R_{0}$ to $R_{1}$ in the process $1\rightarrow2$ and inversely
in $3\rightarrow4$. The system evolves under the time-dependent Hamiltonian
as $\dot{\rho}=-i[H(t),\rho]$. The work done for the finite-time
adiabatic process equals to the change of the internal energy 
\begin{equation}
W(\tau)=\mathrm{Tr}[\rho(\tau)H(\tau)]-\mathrm{Tr}[\rho(0)H(0)],\label{eq:work}
\end{equation}
where $\tau$ is the control time of the adiabatic process. The initial
state $\rho(0)$ is a thermal state, while the final state $\rho(\tau)$
is not necessarily a thermal state. The finite-time adiabatic process
requires more work compared to the quasi-static one. In Ref. \citep{Chen2019},
we rewrite the work as
\begin{equation}
W(\tau)=W^{\mathrm{adi}}+W^{(\mathrm{ex})}(\tau),\label{eq:divide work}
\end{equation}
where $W^{\mathrm{adi}}$ is the work done in the quasi-static adiabatic
process with infinite control time, and $W^{(\mathrm{ex})}(\tau)$
is the extra work for the finite-time adiabatic process.

We have shown that the extra work can be naturally divided into the
mean extra work and the oscillating extra work
\begin{equation}
W^{\mathrm{(ex)}}(\tau)=W^{\mathrm{(mean)}}(\tau)+W^{\mathrm{(osc)}}(\tau).\label{eq:divide extra work}
\end{equation}
The mean extra work decreases monotonously for longer control time
$\tau$, satisfying the $\mathcal{C}/\tau^{2}$ scaling behavior.
The oscillating extra work oscillates around zero with the increasing
control time. For a large and complicated physical system, the oscillating
extra work is usually neglected due to the incommensurable energy
levels of different states \citep{Chen2019}. However, in quantum
systems with simple energy-level structure, the contribution of the
oscillating extra work should be taken into account.

To evaluate the efficiency, one needs to obtain the heat transfer
in the isochoric process $4\rightarrow1$ ($2\rightarrow3$). Since
no work is performed in this process, the heat is determined by the
change of the internal energy. In the process $4\rightarrow1$, the
system absorbs the heat from the hot source $Q_{\mathrm{h}}=\left\langle H\right\rangle _{1}-\left\langle H\right\rangle _{4}>0$,
while the system releases the heat to the cold sink $Q_{\mathrm{c}}=\left\langle H\right\rangle _{3}-\left\langle H\right\rangle _{2}<0$
in the process $2\rightarrow3$. The time consuming of the isochoric
process can be neglected compared to that of the adiabatic process
\citep{Chotorlishvili2016,Abah2018}. For a whole cycle, the net work
is $W_{\mathrm{T}}=Q_{\mathrm{h}}-\left|Q_{\mathrm{c}}\right|$ with
the efficiency $\eta=W_{\mathrm{T}}/Q_{\mathrm{h}}$.

In Ref. \citep{Chen2019}, we have obtained the power
\begin{equation}
P=\frac{W_{\mathrm{T}}^{\mathrm{adi}}-W_{1}^{(\mathrm{ex})}(\tau_{1})-W_{3}^{(\mathrm{ex})}(\tau_{3})}{\tau_{1}+\tau_{3}},\label{eq:power}
\end{equation}
and the efficiency

\begin{equation}
\eta=\frac{W_{\mathrm{T}}^{\mathrm{adi}}-W_{1}^{(\mathrm{ex})}(\tau_{1})-W_{3}^{(\mathrm{ex})}(\tau_{3})}{Q_{\mathrm{h}}^{\mathrm{adi}}-W_{3}^{(\mathrm{ex})}(\tau_{3})}\label{eq:efficiency}
\end{equation}
for the finite-time Otto cycle. Here, $W_{\mathrm{T}}^{\mathrm{adi}}$
and $Q_{\mathrm{h}}^{\mathrm{adi}}$ denote the net work and the heat
absorbed from the hot source in the quasi-static Otto cycle. $W_{1}^{(\mathrm{ex})}(\tau_{1})$
and $W_{3}^{(\mathrm{ex})}(\tau_{3})$ denote the extra work for the
finite-time adiabatic processes $1\rightarrow2$ and $3\rightarrow4$
respectively. For given control time $\tau_{1}$ and $\tau_{3}$,
higher power and efficiency can be achieved by optimizing the protocol
to reduce the extra work $W_{1}^{(\mathrm{ex})}(\tau_{1})$ and $W_{3}^{(\mathrm{ex})}(\tau_{3})$.

In the previous paper, the constraint relation between the efficiency
and the output power is obtained by neglecting the oscillating extra
work for the system with complex energy-level structure. We only consider
the mean part in the extra work $W_{1}^{(\mathrm{ex})}(\tau_{1})\approx\Sigma_{1}/\tau_{1}^{2}$
and $W_{3}^{(\mathrm{ex})}(\tau_{3})\approx\Sigma_{3}/\tau_{3}^{2}$,
and obtain the efficiency at the maximum power as
\begin{equation}
\eta_{\mathrm{EMP}}=\frac{2\eta^{\mathrm{adi}}}{3-\eta^{\mathrm{adi}}/[1+(\Sigma_{1}/\Sigma_{3})^{1/3}]},
\end{equation}
where $\eta^{\mathrm{adi}}$ is the efficiency of the quasi-static
Otto cycle. Yet, such simplification fails for a quantum system with
simple energy-level structure. We will explore the effect of the oscillating
extra work for the simple quantum system in the following section.

\begin{figure}
\includegraphics[width=8cm]{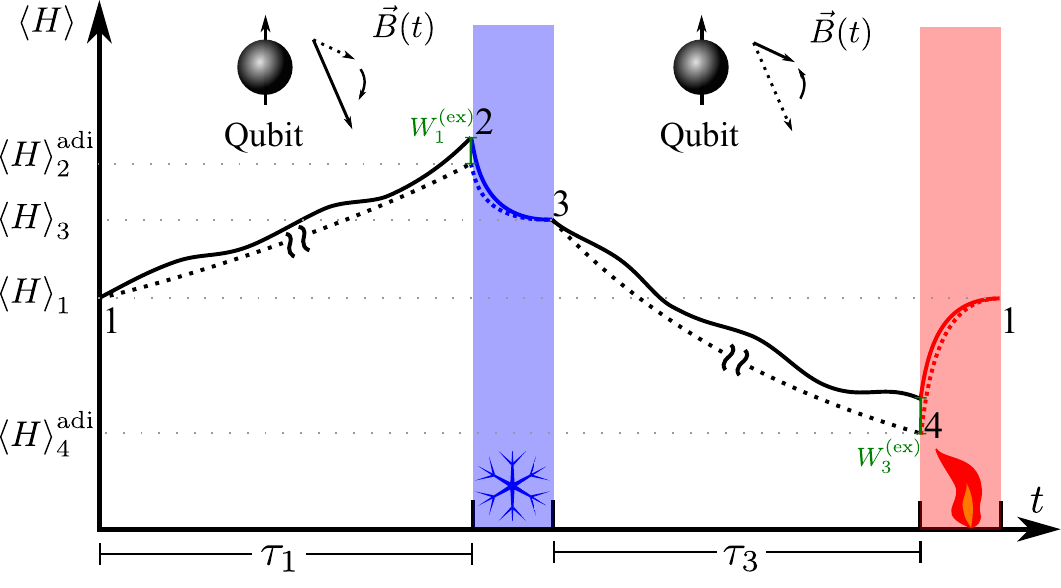}

\caption{\label{fig:two-level-diagram} Finite-time Otto cycle of the two level
system. The magnetic field $\vec{B}(t)$ is modulated in the finite-time
adiabatic process. Similar to Fig. \ref{fig:Finite-time-Otto-cycle.},
the solid line presents the finite-time cycle, with the dashed line
for the quasi-static one plotting for comparison. The time consuming
for the two adiabatic processes is $\tau_{1}$ and $\tau_{3}$, while
the time consuming for the isochoric processes is neglected.}
\end{figure}

\section{Two-level Otto engine\label{sec:Two-level-Otto-heat}}

To show the effect of the oscillating extra work, we start with the
simplest model of two-level system, a spin in a controllable magnetic
field $\vec{B}(t)$. The Hamiltonian of the system reads

\begin{equation}
H=\mu\vec{B}(t)\cdot\vec{\sigma},\label{eq:Hamiltonian}
\end{equation}
with the magnetic moment $\mu$ and the Pauli matrix $\vec{\sigma}$.
We consider the magnetic field is modulated as $\vec{B}(t)=\left(B_{z}(t)+B_{\theta}\cos\theta\right)\vec{e}_{z}+B_{\theta}\sin\theta\vec{e}_{x}$
in the finite-time adiabatic process, where $\theta$ is the angle
of the static magnetic field $B_{\theta}$. With the ratio of the
magnetic field $\lambda=B_{z}(t)/B_{\theta}$, the Hamiltonian of
the two-level system \citep{Alecce2015} is rewritten as 
\begin{equation}
H=\epsilon[(\lambda-\cos\theta)\sigma_{z}+\sin\theta\:\sigma_{x}],
\end{equation}
 by setting $\epsilon=\mu B_{\theta}$ as the unit of the energy.
Here, $\lambda=\lambda(t)$ serves as the tuning parameter $R(t)$
in the finite-time adiabatic process. Figure \ref{fig:two-level-diagram}
shows the finite-time Otto cycle realized on the two-level system.
We present the finite-time cycle with the solid curve, and the quasi-static
cycle with the dashed curve. In the two isochoric processes, the magnetic
field is fixed and the system contacts with the hot source or the
cold sink and reaches equilibrium (the red and blue curve).

To apply the high-order adiabatic approximation, we rewrite the Hamiltonian
under the basis of instantaneous eigenstates \citep{Sun_1988} as
\begin{equation}
H=\epsilon\Lambda(t)\left(\left|e(t)\right\rangle \left\langle e(t)\right|-\left|g(t)\right\rangle \left\langle g(t)\right|\right),\label{eq:Hamiltonian-two-level}
\end{equation}
where the instantaneous eigen-energy $\epsilon\Lambda(t)$ is determined
by $\lambda(t)$ as

\begin{equation}
\Lambda(t)=\sqrt{\lambda^{2}-2\lambda\cos\theta+1}.
\end{equation}
The instantaneous ground state is 
\begin{align}
\left|g(t)\right\rangle  & =1/N_{1}\left(\begin{array}{c}
\lambda-\cos\theta-\Lambda\\
\sin\theta
\end{array}\right),
\end{align}
and the instantaneous excited state is

\begin{equation}
\left|e(t)\right\rangle =1/N_{2}\left(\begin{array}{c}
\lambda-\cos\theta+\Lambda\\
\sin\theta
\end{array}\right),
\end{equation}
where $N_{1}=[2\Lambda(\Lambda-\lambda+\cos\theta)]^{1/2}$ and $N_{2}=[2\Lambda(\Lambda+\lambda-\cos\theta)]^{1/2}$
are the normalized factors.

The initial state is a thermal state $\rho(0)=p_{g}\left|g(0)\right\rangle \left\langle g(0)\right|+p_{e}\left|e(0)\right\rangle \left\langle e(0)\right|$,
where the distribution is $p_{g}=1-p_{e}=1/[1+\exp(-2\beta\epsilon\Lambda(0)]$
with the inverse temperature $\beta$. The density matrix at any time
$t\in[0,\tau]$ is $\rho(t)=p_{g}\left|\psi_{g}(t)\right\rangle \left\langle \psi_{g}(t)\right|+p_{e}\left|\psi_{e}(t)\right\rangle \left\langle \psi_{e}(t)\right|$,
where the state $\left|\psi_{n}(t)\right\rangle ,\,n=e,g$ obeys the
Schrodinger equation 
\begin{equation}
i\partial_{t}\left|\psi_{n}(t)\right\rangle =H(t)\left|\psi_{n}(t)\right\rangle ,\label{eq:Schrodinger Eq}
\end{equation}
with the initial condition $\left|\psi_{n}(0)\right\rangle =\left|n(0)\right\rangle $.
We express the state under the basis of the instantaneous eigenstates
\begin{equation}
\left|\psi_{n}(t)\right\rangle =c_{ng}(t)e^{i\phi(t)}\left|g(t)\right\rangle +c_{ne}(t)e^{-i\phi(t)}\left|e(t)\right\rangle \label{eq:psi_n}
\end{equation}
with the dynamical phase $\phi(t)=\epsilon\int_{0}^{t}\Lambda(t^{\prime})dt^{\prime}$.
The Schrodinger equation by Eq. (\ref{eq:Schrodinger Eq}) gives the
differential equations

\begin{align}
\text{\ensuremath{\dot{c}}}_{ng} & =e^{-2i\phi}\frac{\sin\left|\theta\right|\dot{\lambda}}{2\Lambda^{2}}c_{ne},\label{eq:cng}
\end{align}
and

\begin{equation}
\dot{c}_{ne}=-e^{2i\phi}\frac{\sin\left|\theta\right|\dot{\lambda}}{2\Lambda^{2}}c_{ng}.\label{eq:cne}
\end{equation}

We consider a given protocol $\tilde{\lambda}(s)=\lambda(s\tau)$
with adjustable control time $\tau$, where $s=t/\tau\in[0,1]$ denotes
the rescaled time parameter. The internal energy at the end of the
adiabatic process is $\left\langle H(\tau)\right\rangle =(p_{e}-p_{g})[1-2\left|c_{eg}\left(\tau\right)\right|^{2}]\epsilon\tilde{\Lambda}(1)$
with the notation $\tilde{\Lambda}(s)=\Lambda(s\tau)$. Together with
Eqs. (\ref{eq:work}) and (\ref{eq:divide work}), we obtain the quasi-static
work
\begin{equation}
W^{\mathrm{adi}}=-\epsilon\tanh[\beta\epsilon\tilde{\Lambda}(0)][\tilde{\Lambda}(1)-\tilde{\Lambda}(0)],
\end{equation}
and the extra work for the finite-time adiabatic process
\begin{equation}
W^{\mathrm{(ex)}}(\tau)=2\epsilon\tilde{\Lambda}(1)\tanh[\beta\epsilon\tilde{\Lambda}(0)]\left|c_{ge}\left(\tau\right)\right|^{2}.\label{eq:extra work exact}
\end{equation}

At long control time limit, the first-order adiabatic approximation
gives the asymptotic amplitude for Eq. (\ref{eq:cne})

\begin{align}
c_{ge}^{[1]}\left(\tau\right) & =\frac{i\sin\left|\theta\right|}{4\epsilon\tau}\left(\frac{\tilde{\lambda}^{\prime}(1)}{\tilde{\Lambda}(1)^{3}}e^{2i\tau\tilde{\phi}\left(1\right)}-\frac{\tilde{\lambda}^{\prime}(0)}{\tilde{\Lambda}(0)^{3}}\right),\label{eq:cge=00005B1=00005D-1-1}
\end{align}
where the dynamical phase is rewritten as $\tilde{\phi}(s)=\epsilon\int_{0}^{s}\tilde{\Lambda}(s^{\prime})ds^{\prime}$,
and $\tilde{\lambda}^{\prime}(s)=\mathrm{d}\tilde{\lambda}(s)/\mathrm{d}s$
denotes the derivative of $\tilde{\lambda}(s)$. The derivation of
Eq. (\ref{eq:cge=00005B1=00005D-1-1}) is attached in Appendix \ref{sec:two-level system appendix}.
Substituting Eq. (\ref{eq:cge=00005B1=00005D-1-1}) into the extra
work by Eq. (\ref{eq:extra work exact}), the asymptotic extra work
is naturally divided into two parts according to Eq. (\ref{eq:divide extra work}),
the mean extra work

\begin{equation}
W^{\mathrm{(mean)}}=\frac{\sin^{2}\theta\tilde{\Lambda}(1)}{8\epsilon\tau^{2}}\left(\frac{\tilde{\lambda^{\prime}}(1)^{2}}{\tilde{\Lambda}(1)^{6}}+\frac{\tilde{\lambda^{\prime}}(0)^{2}}{\tilde{\Lambda}(0)^{6}}\right)\tanh[\beta\epsilon\tilde{\Lambda}(0)],\label{eq:mean}
\end{equation}
and the oscillating extra work

\begin{equation}
W^{\mathrm{(osc)}}=-\frac{\sin^{2}\theta}{4\epsilon\tau^{2}}\frac{\tilde{\lambda^{\prime}}(1)\tilde{\lambda^{\prime}}(0)\cos[2\tau\tilde{\phi}(1)]}{\tilde{\Lambda}(1)^{2}\tilde{\Lambda}(0)^{3}}\tanh[\beta\epsilon\tilde{\Lambda}(0)].\label{eq:osc work}
\end{equation}

To obtain the efficiency and the power for the finite-time Otto cycle,
we need the net work $W_{\mathrm{T}}^{\mathrm{adi}}$ and the heat
absorbed $Q_{\mathrm{h}}^{\mathrm{adi}}$ in the quasi-static Otto
cycle with the infinite control time $\tau\rightarrow\infty$. The
magnetic field $B_{z}(t)$ is modulated from $B_{0}$ to $B_{1}$
in the adiabatic process $1\rightarrow2$, with the corresponding
parameter $\lambda_{0}$ and $\lambda_{1}$ at the initial and final
time. Since the population on the excited state remains unchanged
during the quasi-static adiabatic processes \citep{Quan_2007}, the
internal energy of the four states follows immediately as $\left\langle H\right\rangle _{1}=-E_{0}\tanh(\beta_{\mathrm{h}}E_{0})$,
$\left\langle H\right\rangle _{2}^{\mathrm{adi}}=-E_{1}\tanh(\beta_{\mathrm{h}}E_{0})$,
$\left\langle H\right\rangle _{3}=-E_{1}\tanh(\beta_{\mathrm{c}}E_{1})$
and $\left\langle H\right\rangle _{4}^{\mathrm{adi}}=-E_{0}\tanh(\beta_{\mathrm{c}}E_{1})$.
Here, $\beta_{l}=1/k_{\mathrm{B}}T_{l}$ is the inverse temperature
for the hot source ($l=\mathrm{h}$) and the cold sink ($l=\mathrm{c}$),
and $E_{j}=\epsilon\sqrt{\lambda_{j}^{2}-2\lambda_{j}\cos\theta+1},\,j=0,1$
gives the abbreviation of the eigen-energy. The net work of the quasi-static
Otto cycle is 
\begin{align}
W_{\mathrm{T}}^{\mathrm{adi}} & =(E_{0}-E_{1})[\tanh(\beta_{\mathrm{c}}E_{1})-\tanh(\beta_{\mathrm{h}}E_{0})]\label{eq:adi work}
\end{align}
The heat absorbed from the hot source is

\begin{equation}
Q_{\mathrm{h}}^{\mathrm{adi}}=E_{0}[\tanh(\beta_{\mathrm{c}}E_{1})-\tanh(\beta_{\mathrm{h}}E_{0})].\label{eq:adi heat}
\end{equation}
The efficiency of the quasi-static Otto cycle is $\eta^{\mathrm{adi}}=1-E_{1}/E_{0}$
\citep{Quan_2007}. For the finite-time Otto cycle, the power and
the efficiency are obtained by substituting the extra work by Eq.
(\ref{eq:extra work exact}) and the quasi-static net work and heat
by Eqs. (\ref{eq:adi work}) and (\ref{eq:adi heat}) into Eqs. (\ref{eq:power})
and (\ref{eq:efficiency}), respectively.

\begin{figure}
\includegraphics[width=8cm]{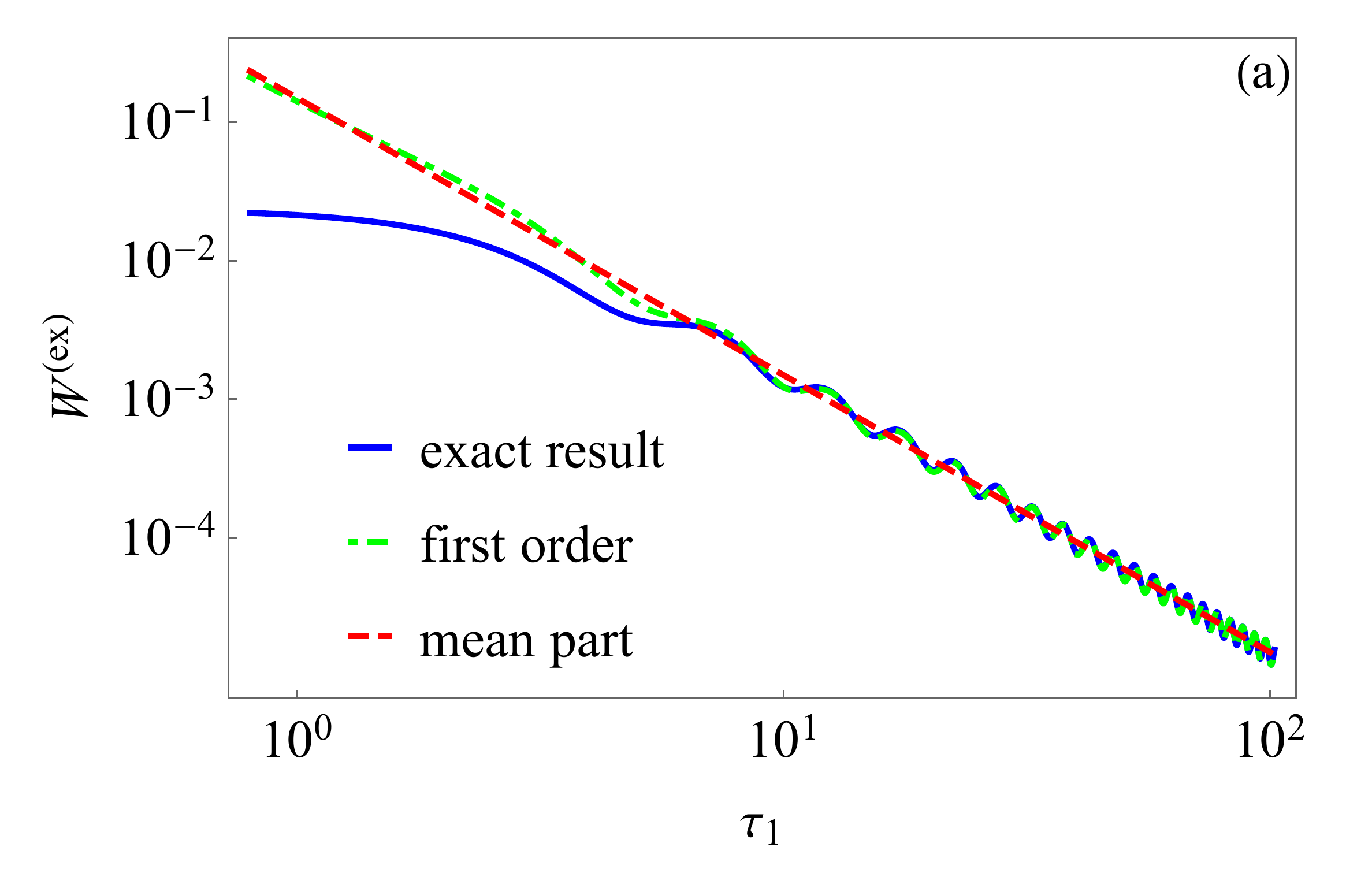}

\includegraphics[width=8cm]{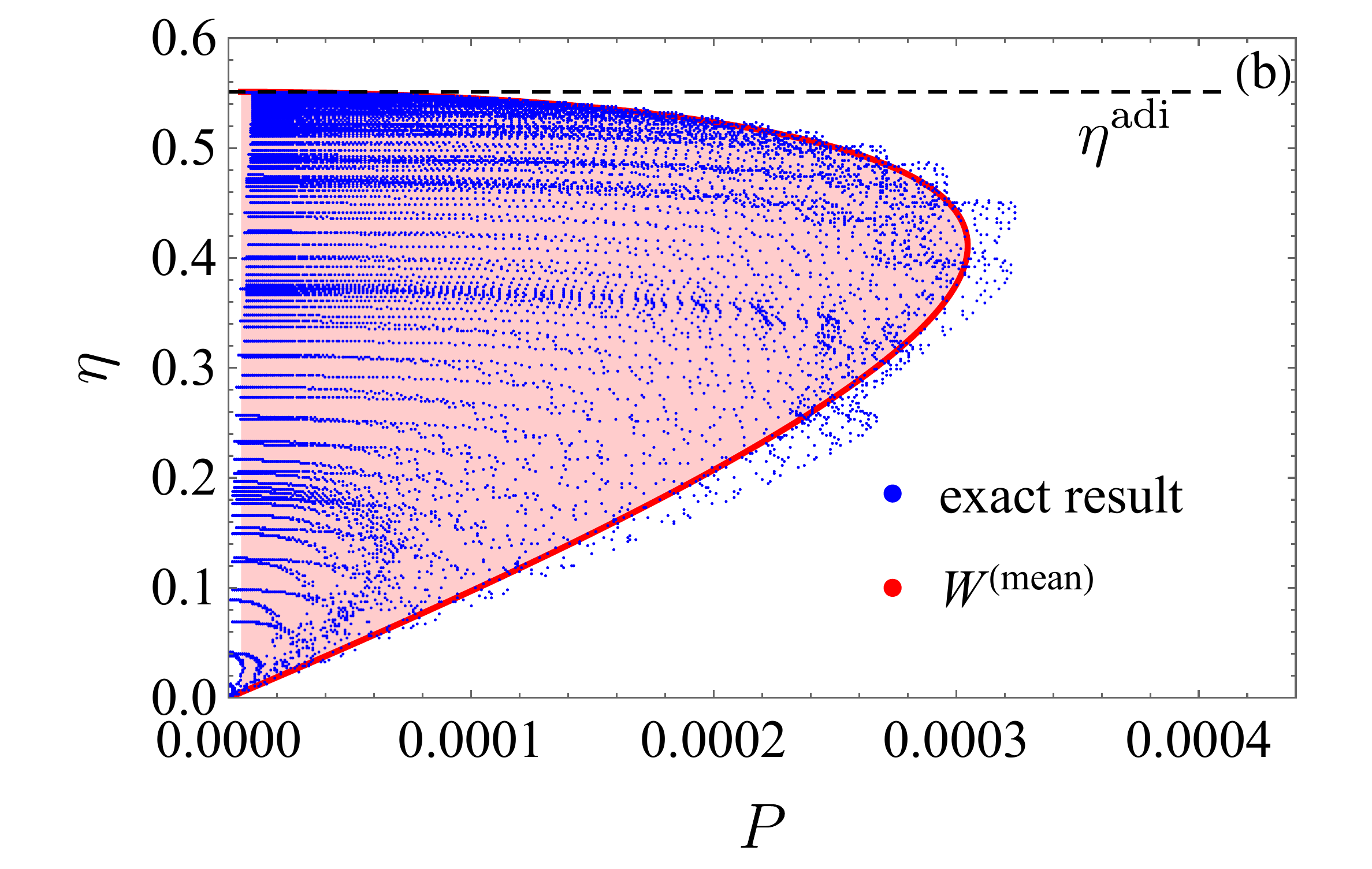}

\caption{\label{fig:comparing}(a) The extra work for the adiabatic process
$1\rightarrow2$ with the linear protocol $\tilde{\lambda}_{l}$.
The blue-solid curve, the green-dashdotted curve and the red-dashed
line present the exact numerical result, the first-order adiabatic
result, and the mean extra work, respectively, while the horizontal
black-dashed line presents the quasi-static efficiency. The parameters
are chosen as $\lambda_{0}=0.1$ and $\lambda_{1}=0.8$ with $\theta=0.4$
and $\epsilon=1$, with the temperatures $T_{\mathrm{h}}=5$ and $T_{\mathrm{c}}=2$.
(b) the reachable power and efficiency for the finite-time two-level
Otto engine. The red area only accounts for the mean extra work, while
the blue dots present the exact numerical result.}
\end{figure}

We compare the asymptotic extra work by Eqs (\ref{eq:mean}) and (\ref{eq:osc work})
with the exact numerical result in Fig. \ref{fig:comparing}(a). The
exact numerical result is obtained by numerically solving Eq. (\ref{eq:cng})
and (\ref{eq:cne}). We choose the parameters $\theta=0.4,\,\epsilon=1,\,\lambda_{0}=0.1,\,\lambda_{1}=0.8$,
and set temperatures for the hot source and cold sink as $k_{B}T_{\mathrm{h}}=5$
and $k_{B}T_{\mathrm{c}}=2$. We first adopt the linear protocol $\tilde{\lambda}_{l}(t/\tau_{1})=\tilde{\lambda}(0)+[\tilde{\lambda}(1)-\tilde{\lambda}(0)]t/\tau_{1}$.
Figure \ref{fig:comparing}(a) shows the extra work for the adiabatic
process $1\rightarrow2$ with different control time $\tau_{1}$,
where the initial and the final tuning parameter are $\tilde{\lambda}(0)=\lambda_{0}$
and $\tilde{\lambda}(1)=\lambda_{1}$. The extra work (the blue curve)
decreases with oscillation with the increasing control time, satisfying
the $\mathcal{C}/\tau^{2}$ scaling (the red-dashed line). The asymptotic
extra work from the first-order adiabatic approximation (the green-dashdotted
curve) matches with the exact numerical result (the blue curve) at
long control time.

We evaluate the performance of the finite-time Otto engine by modulating
the control time $\tau_{1}$ and $\tau_{3}$ for the finite-time adiabatic
processes. Figure \ref{fig:comparing}(b) illustrates the constraint
relation between efficiency and power. The red area presents the result
with the mean extra work $W^{\mathrm{(mean)}}(\tau)$, where the oscillating
extra work is neglected. The blue dots present the exact result by
numerically calculating the extra work $W_{1}^{(\mathrm{ex})}(\tau_{1})$
and $W_{3}^{(\mathrm{ex})}(\tau_{3})$ in finite time. The oscillation
of the extra work leads to a jagged edge in the constraint relation,
and can be utilized to achieve larger maximum power.

To attain high power, we should reduce the extra work in the finite-time
adiabatic processes at the given control time $\tau_{1}$ or $\tau_{3}$.
The extra work by Eqs (\ref{eq:mean}) and (\ref{eq:osc work}) approaches
zero at the specific control time $\tau=n\pi/\tilde{\phi}(1),\,n=1,2,...$
with the condition $\tilde{\lambda^{\prime}}(1)/[\tilde{\Lambda}(1)]^{3}=\tilde{\lambda^{\prime}}(0)/[\tilde{\Lambda}(0)]^{3}$.
We design such special protocol $\tilde{\lambda}_{s}(s)$ to satisfy
this condition, determined by the implicit equation

\begin{equation}
s=\frac{\frac{\tilde{\lambda}(s)-\cos\theta}{\tilde{\Lambda}(s)}-\frac{\tilde{\lambda}(0)-\cos\theta}{\tilde{\Lambda}(0)}}{\frac{\tilde{\lambda}(1)-\cos\theta}{\tilde{\Lambda}(1)}-\frac{\tilde{\lambda}(0)-\cos\theta}{\tilde{\Lambda}(0)}}.\label{eq:specialprotocol}
\end{equation}
 By adopting this special protocol for the finite-time adiabatic processes,
the efficiency of the Otto cycle approaches the quasi-static one $\eta^{\mathrm{adi}}=W_{\mathrm{T}}^{\mathrm{adi}}/Q_{\mathrm{h}}^{\mathrm{adi}}$
with finite output power.

\begin{figure}
\includegraphics[width=8cm]{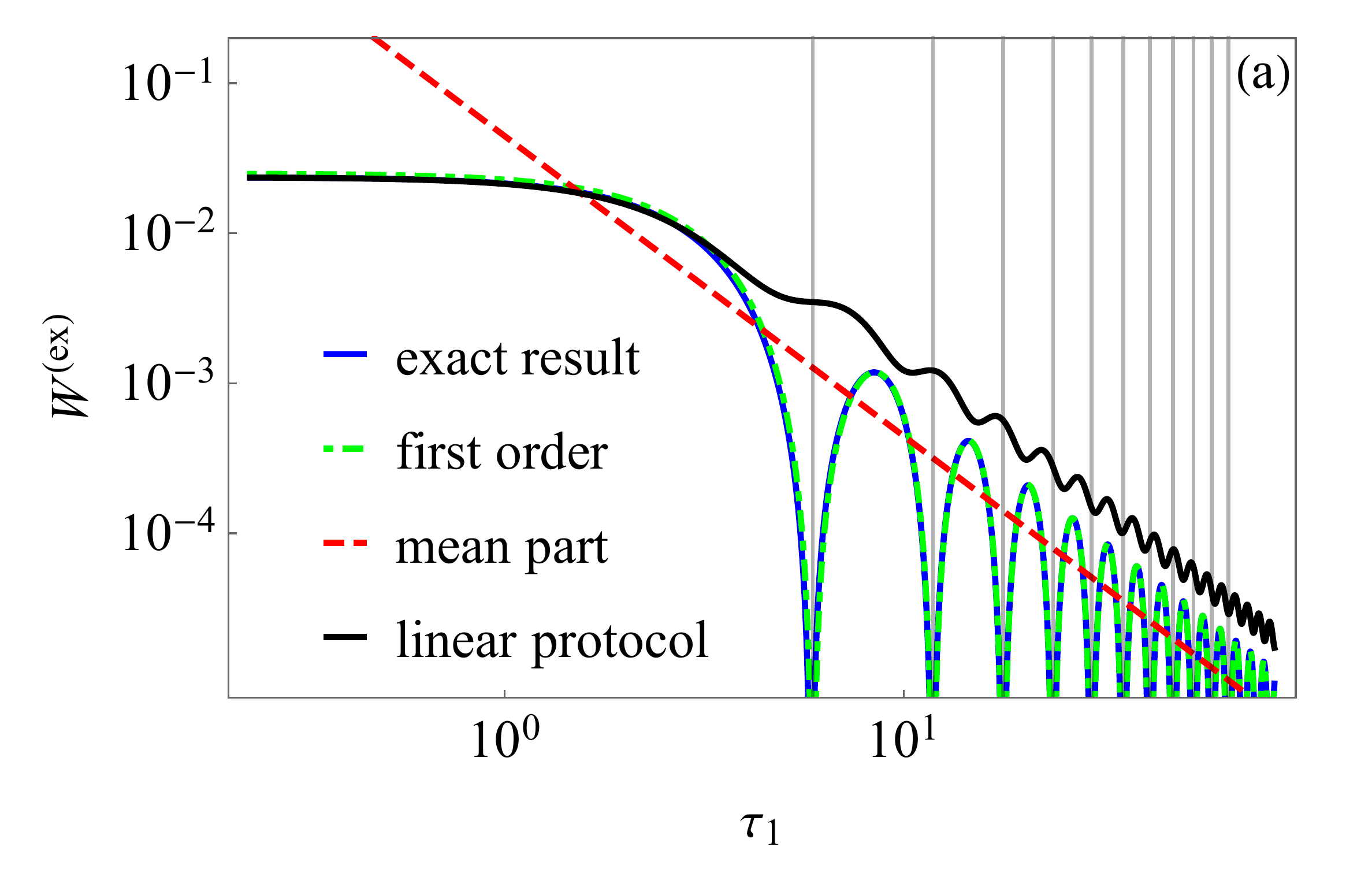}

\includegraphics[width=8cm]{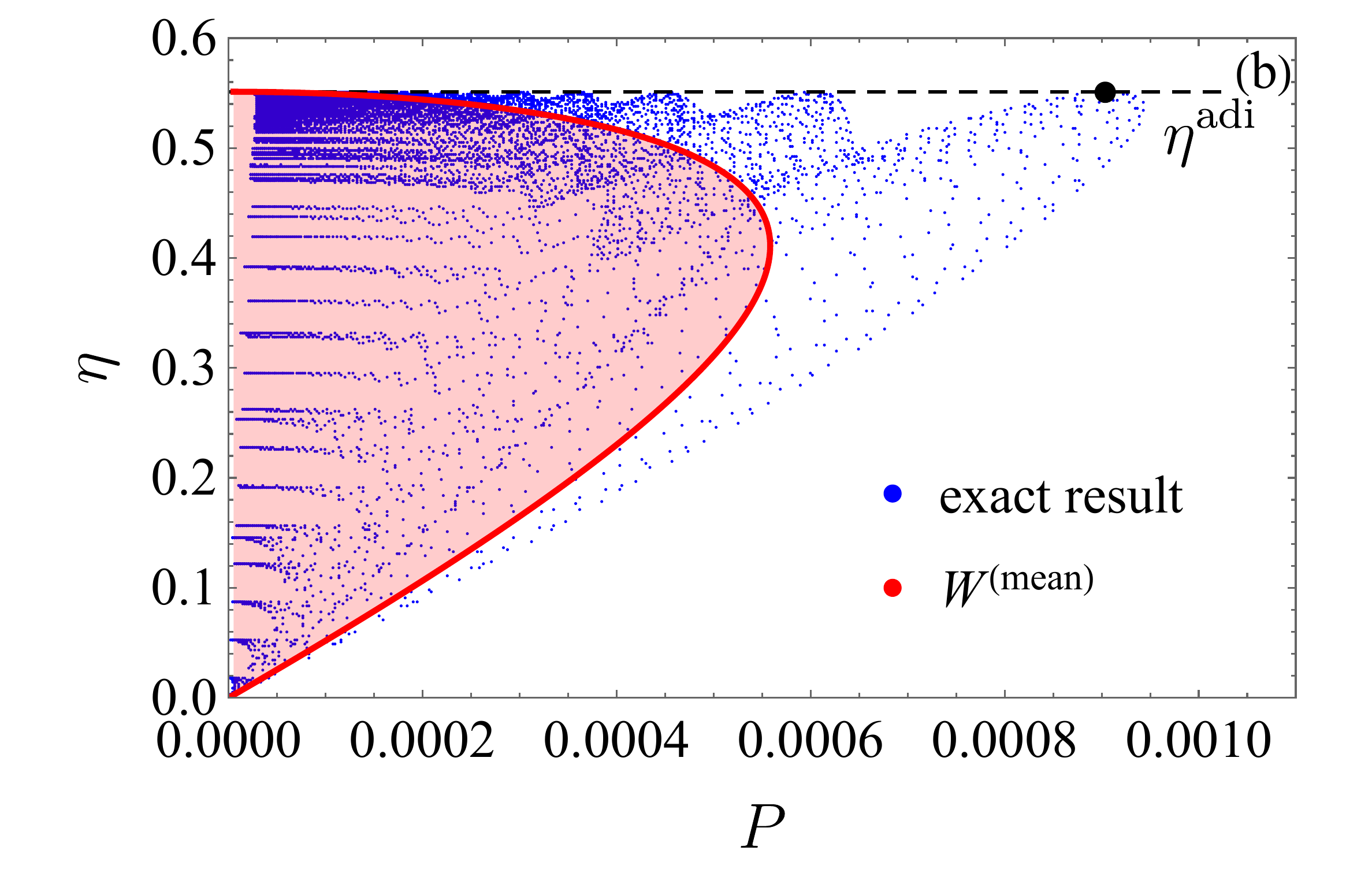}

\caption{\label{fig:comparing2specialprotocol}(a) The extra work for the adiabatic
process $1\rightarrow2$ with the special protocol $\tilde{\lambda}_{s}$
by Eq (\ref{eq:specialprotocol}). The black curve presents the exact
extra work in the previous linear protocol for comparing. The vertical
gray-dashed line shows the extra work approaches zero at the specific
control time. (b) The reachable power and efficiency with the special
protocol. The same parameters are chosen in Fig. \ref{fig:comparing}.}
\end{figure}

Figure \ref{fig:comparing2specialprotocol}(a) presents the first-order
adiabatic extra work (the green-dashdotted curve), the mean extra
work (the red-dashed line) and the exact one (the blue-solid curve)
for the designed protocol by Eq. (\ref{eq:specialprotocol}). The
extra work for the linear protocol (the black-solid curve) is plotted
for comparison. The dynamical phase of the special protocol is $\tilde{\phi}(1)=0.531$,
obtained by Eq. (\ref{eq:phitheta}) in the Appendix \ref{sec:two-level system appendix}.
Hence, the extra work approaches zero at the specific control time
$\tau=n\pi/\tilde{\phi}(1)=n\times5.92,\,n=1,2,...$, shown as the
vertical gray-dashed line.

Figure \ref{fig:comparing2specialprotocol}(b) presents the constraint
relation between the efficiency and the power for the special protocol.
When the control time of the two adiabatic processes is chosen as
the specific control time $\tau=n\pi/\tilde{\phi}(1)$, the efficiency
approaches to the quasi-static efficiency $\eta^{\mathrm{adi}}=0.551$
(the horizontal black-dashed line). For the specific control time
$\tau_{1}=\tau_{3}=5.92$, the heat engine gains large power with
the quasi-static efficiency, marked with the black point. Compared
to the linear protocol, the quantum Otto engine with the special protocol
attains larger maximum power and the higher efficiency.

By optimizing the control time of the quantum adiabatic processes,
the oscillating extra work can be utilized to improve the maximum
power and the efficiency for the finite-time Otto engine. In the next
section, we continue to study the effect of similar oscillation of
the extra work on the Otto cycle with quantum harmonic oscillator.

\section{Quantum Harmonic Otto Engine\label{sec:Quantum-Harmonic-Otto}}

Another system with simple energy level structure is the quantum harmonic
oscillator, which has been widely studied as a prototype of the quantum
Otto engine \citep{Insinga2016,Kosloff2017,Deffner2018}. The technique
of shortcut to adiabaticity has been applied to ameliorate the quantum
harmonic Otto engine \citep{Deng2013,Chen2010PhysRevLett104_63002,Abah2018,Cakmak2019}.
Here, we consider a generic finite-time adiabatic process described
by the time-dependent Hamiltonian
\begin{equation}
H=-\frac{1}{2M}\frac{\partial^{2}}{\partial x^{2}}+\frac{1}{2}M\omega^{2}x^{2}.
\end{equation}
The frequency $\omega=\omega(t),\,t\in[0,\tau]$ serves as the tuning
parameter in the finite-time adiabatic process. The wave function
of the instantaneous eigenstate is
\begin{equation}
\left\langle x\left|n(t)\right.\right\rangle =N_{n}\exp(-\frac{1}{2}M\omega x^{2})H_{n}(\sqrt{M\omega}x),\label{eq:eigenstate}
\end{equation}
with the corresponding instantaneous eigen-energy $E_{n}(t)=(n+1/2)\omega(t)$.
$H_{n}(\xi)=(-1)^{n}\exp(\xi^{2})\partial^{n}/\partial\xi^{n}[\exp(-\xi^{2})]$
denotes the Hermite polynomial with the order $n$, and $N_{n}=\left(\sqrt{M\omega}/\sqrt{\pi}2^{n}n!\right)^{1/2}$
is the normalized factor.

\begin{figure}
\includegraphics[width=8cm]{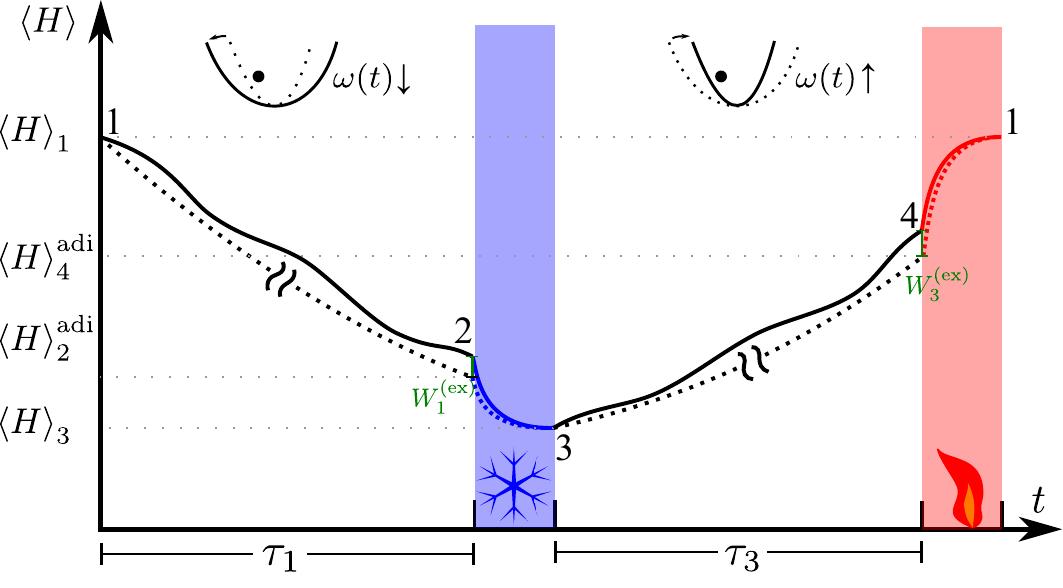}

\caption{\label{fig:quantum harmonic} Finite-time Otto cycle of the quantum
harmonic oscillator. The frequency $\omega(t)$ of the oscillator
is modulated in the finite-time adiabatic process.}
\end{figure}

Figure \ref{fig:quantum harmonic} illustrates the finite-time quantum
harmonic Otto cycle. Similar to the Otto cycle of the two-level system,
the work is performed in two adiabatic processes $1\rightarrow2$
and $3\rightarrow4$, while the system exchanges the heat with the
hot source (cold sink) and reaches equilibrium in the isochoric process
$4\rightarrow1$ ($2\rightarrow3$).

In the two adiabatic process $1\rightarrow2$ and $3\rightarrow4$,
the initial state is the thermal state $\rho(0)=\sum_{n=0}^{\infty}p_{n}\left|n(0)\right\rangle \left\langle n(0)\right|$
with the distribution $p_{n}=2\sinh(\beta\omega(0)/2)\exp[-\beta(n+1/2)\omega(0)].$
The density matrix at any time $t$ during the evolution is $\rho(t)=\sum_{n=0}^{\infty}p_{n}\left|\psi_{n}(t)\right\rangle \left\langle \psi_{n}(t)\right|$.
Here, the state $\left|\psi_{n}(t)\right\rangle ,\,n=0,1,2...$ obeys
the Schrodinger equation $i\partial_{t}\left|\psi_{n}(t)\right\rangle =H(t)\left|\psi_{n}(t)\right\rangle ,$with
the initial condition $\left|\psi_{n}(0)\right\rangle =\left|n(0)\right\rangle $.
Similar to Eq. (\ref{eq:Hamiltonian-two-level}), we rewrite the state
under the instantaneous diagonal basis
\begin{equation}
\left|\psi_{n}(t)\right\rangle =\sum_{m}c_{nm}(t)e^{-i(m+\frac{1}{2})\varphi(t)}\left|n(t)\right\rangle ,
\end{equation}
with the dynamical phase $\varphi(t)=\int_{0}^{t}\omega(t^{\prime})dt^{\prime}$.
The differential equation of $c_{nm}(t)$ is obtained in Appendix
\ref{sec:appendix Time-dependent-Harmonic-Oscillat}.

To evaluate the finite-time effect of the quantum adiabatic process,
we consider a protocol $\tilde{\omega}(s)=\omega(s\tau)$ with adjustable
control time $\tau$, with the instantaneous energy as $\tilde{E}_{n}(s)=E_{n}(s\tau)$.
We rewrite the work into the quasi-static work and the extra work
as Eq. (\ref{eq:divide work}). The quasi-static work with infinite
control time $\tau\rightarrow\infty$ is
\begin{equation}
W^{\mathrm{adi}}=\sum_{n=0}^{\infty}p_{n}\left[\tilde{E}_{n}(1)-\tilde{E}_{n}(0)\right],
\end{equation}
while extra work in finite-time adiabatic process is

\begin{equation}
W^{\mathrm{(ex)}}(\tau)=\sum_{n,m=0}^{\infty}p_{n}\left|c_{nm}(\tau)\right|^{2}\left[\tilde{E}_{m}(1)-\tilde{E}_{n}(1)\right].\label{eq:extra work harmonic}
\end{equation}

Similar to Eq. (\ref{eq:cge=00005B1=00005D-1-1}), the asymptotic
amplitude at long control time is given by the first-order adiabatic
approximation

\begin{align}
c_{n,n+2}^{[1]}(\tau) & =-i\frac{\sqrt{(n+1)(n+2)}}{8\tau}\left[\frac{\tilde{\omega}^{\prime}(1)}{\tilde{\omega}(1)^{2}}e^{2i\tau\tilde{\varphi}\left(1\right)}-\frac{\tilde{\omega}^{\prime}(0)}{\tilde{\omega}(0)^{2}}\right],\label{eq:cnn+229}
\end{align}
and

\begin{equation}
c_{n,n-2}^{[1]}(\tau)=-i\frac{\sqrt{n(n-1)}}{8\tau}\left[\frac{\tilde{\omega}^{\prime}(1)}{\tilde{\omega}(1)^{2}}e^{-2i\tau\tilde{\varphi}\left(1\right)}-\frac{\tilde{\omega}^{\prime}(0)}{\tilde{\omega}(0)^{2}}\right],\label{eq:cnn-230}
\end{equation}
where the dynamical phase factor is $\tilde{\varphi}(1)=\int_{0}^{1}\tilde{\omega}(s)ds$.
The derivation of Eqs. (\ref{eq:cnn+229}) and (\ref{eq:cnn-230})
is given in Appendix \ref{sec:appendix Time-dependent-Harmonic-Oscillat}.
The terms $c_{nm}^{[1]}(\tau),\,m\ne n,\,n\pm2$ are all zero in the
first-order adiabatic approximation. According to Eq. (\ref{eq:divide extra work}),
the asymptotic extra work at long control time by Eq. (\ref{eq:extra work harmonic})
is divided into the mean one
\begin{equation}
W^{\mathrm{(mean)}}(\tau)=\frac{\tilde{\omega}(1)}{8\tau^{2}}\left[\frac{\tilde{\omega}^{\prime}(0)^{2}}{\tilde{\omega}(0)^{4}}+\frac{\tilde{\omega}^{\prime}(1)^{2}}{\tilde{\omega}(1)^{4}}\right]\sum_{n=0}^{\infty}(n+\frac{1}{2})p_{n},\label{eq:mean extra work oscillator}
\end{equation}
and the oscillating one
\begin{equation}
W^{\mathrm{(osc)}}(\tau)=-\frac{\cos\left[2\tau\tilde{\varphi}(1)\right]}{4\tau^{2}}\frac{\omega^{\prime}(0)\tilde{\omega}^{\prime}(1)}{\omega(0)^{2}\tilde{\omega}(1)}\sum_{n=0}^{\infty}(n+\frac{1}{2})p_{n}.\label{eq:oscillating extra work}
\end{equation}
The exact result of the extra work is obtained from the numerical
calculation of the non-adiabatic factor with an auxiliary differential
equation \citep{Chen2010PhysRevLett104_63002,Abah2012}. The detail
is shown in Appendix \ref{sec:appendix Time-dependent-Harmonic-Oscillat}.

We calculate the net work $W_{\mathrm{T}}^{\mathrm{adi}}$ and the
heat $Q_{\mathrm{h}}^{\mathrm{adi}}$ for the quasi-static Otto cycle.
Since the population on each state remains unchanged during the quantum
adiabatic processes, the internal energy of the four states follows
as $\left\langle H\right\rangle _{1}=\coth\left(\beta_{h}\omega_{0}/2\right)\omega_{0}/2$,
$\left\langle H\right\rangle _{2}^{\mathrm{adi}}=\coth\left(\beta_{h}\omega_{0}/2\right)\omega_{1}/2$,
$\left\langle H\right\rangle _{3}=\coth\left(\beta_{c}\omega_{1}/2\right)\omega_{1}/2$,
and $\left\langle H\right\rangle _{4}^{\mathrm{adi}}=\coth\left(\beta_{c}\omega_{1}/2\right)\omega_{0}/2$.
The net work of the quasi-static cycle is
\begin{align}
W_{\mathrm{T}}^{\mathrm{adi}} & =\frac{\omega_{0}-\omega_{1}}{2}\left[\coth\left(\frac{\beta_{\mathrm{h}}\omega_{0}}{2}\right)-\coth\left(\frac{\beta_{\mathrm{c}}\omega_{1}}{2}\right)\right].
\end{align}
The heat absorbed from the hot source is

\begin{equation}
Q_{\mathrm{h}}^{\mathrm{adi}}=\omega_{0}\left[\coth\left(\frac{\beta_{\mathrm{h}}\omega_{0}}{2}\right)-\coth\left(\frac{\beta_{\mathrm{c}}\omega_{1}}{2}\right)\right].
\end{equation}

We first adopt the linear protocol $\tilde{\omega}_{l}(s)=\tilde{\omega}(0)+[\tilde{\omega}(1)-\tilde{\omega}(0)]s$
for the finite-time adiabatic process. We set the parameters as $\omega_{0}=2,\,\omega_{1}=1,$
and $M=1$, and choose the temperature for the hot source and cold
sink as $T_{\mathrm{h}}=5$ and $T_{\mathrm{c}}=2$ respectively.

\begin{figure}
\includegraphics[width=8cm]{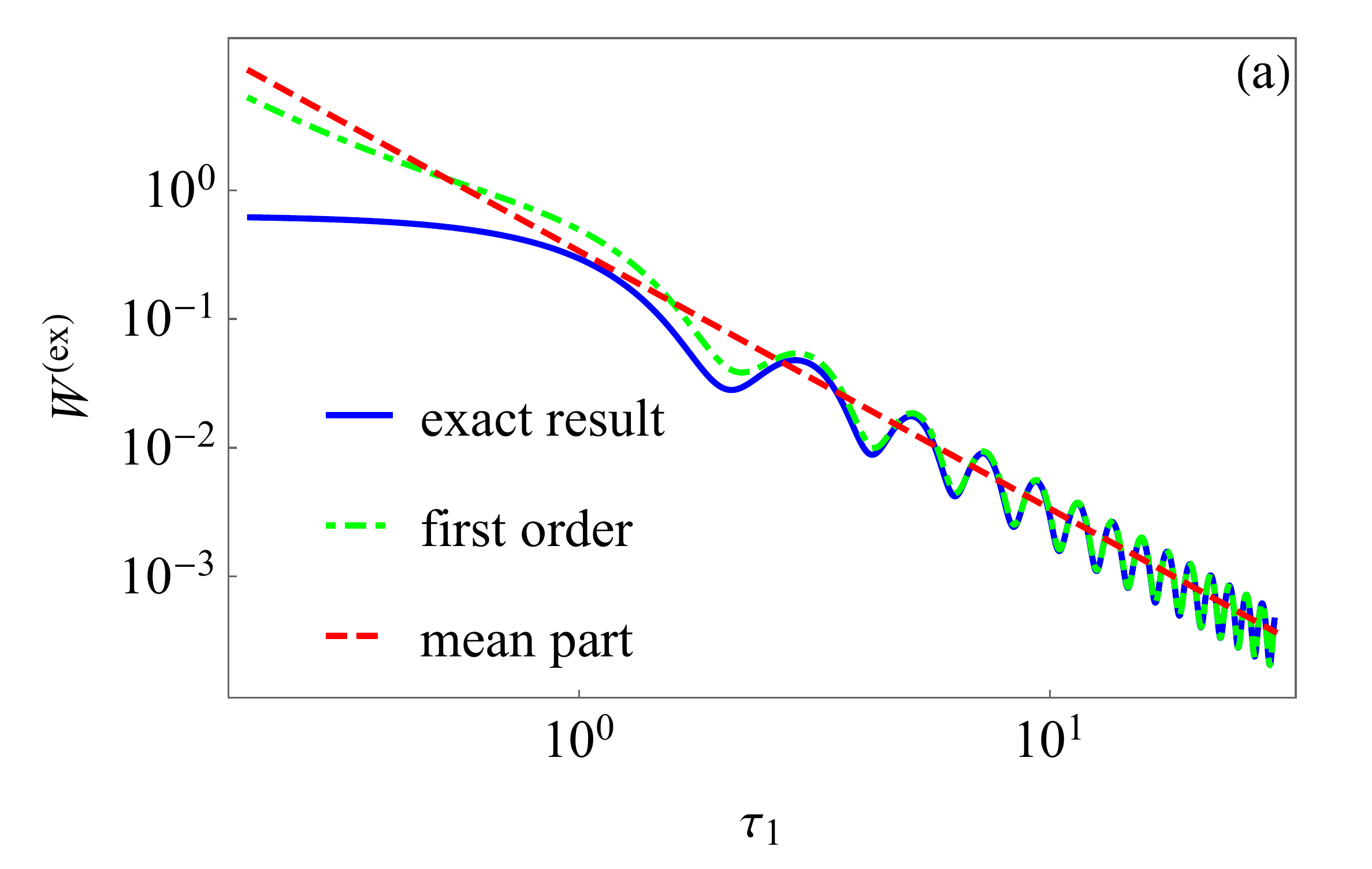}

\includegraphics[width=8cm]{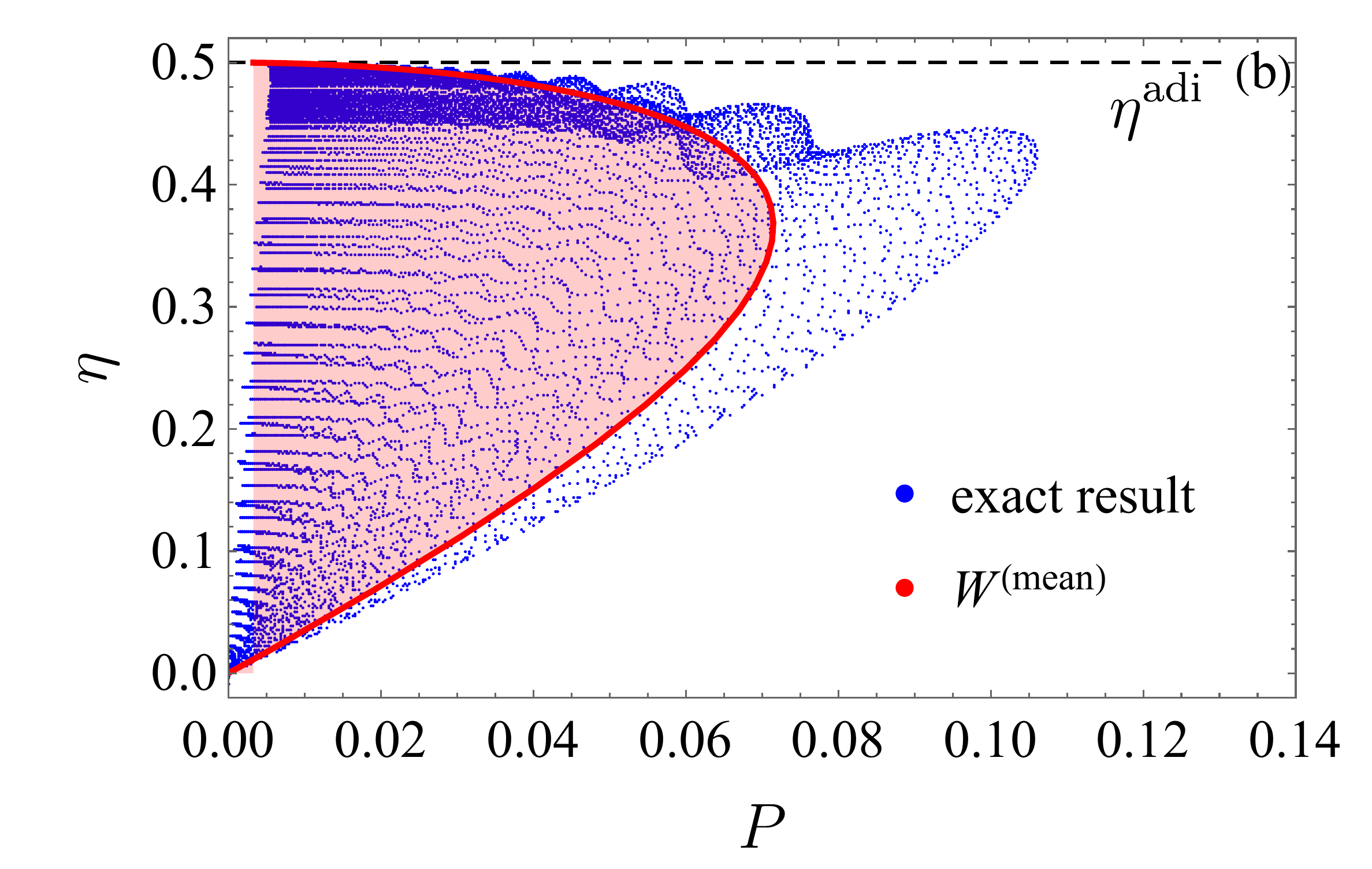}

\caption{\label{fig:comparing-harmonic}(a) The extra work in the adiabatic
process $1\rightarrow2$ with the linear protocol $\tilde{\omega}_{l}$.
The parameter is chosen as $\omega_{0}=2,\,\omega_{1}=1,$ and $M=1$,
with the temperatures $T_{\mathrm{h}}=5$ and $T_{\mathrm{c}}=2$.
(b) the reachable power and efficiency for the quantum harmonic Otto
engine.}
\end{figure}

In Fig. \ref{fig:comparing-harmonic}(a), we compare the first-order
result of extra work with the exact numerical result for the adiabatic
process $1\rightarrow2$ with $\tilde{\omega}(0)=\omega_{0}$ and
$\tilde{\omega}(1)=\omega_{1}$. The first-order adiabatic result
(the green-dashdotted curve) matches with the exact numerical result
(the blue curve) at long control time. The extra work decreases with
oscillation when increasing the control time $\tau_{1}$, retaining
the quantum adiabatic limit with infinite control time. Neglecting
the oscillation, the extra work satisfies the $\mathcal{C}/\tau^{2}$
scaling (the red dashed line). Figure \ref{fig:comparing-harmonic}(b)
shows the constraint relation between the efficiency and the power
for the finite-time quantum harmonic Otto engine. The results are
similar to the two-level Otto engine: the oscillating extra work can
be utilized to obtain higher maximum power with higher efficiency.

To reduce the extra work in the finite-time adiabatic process at the
given control time $\tau$, we consider a special protocol \citep{Deng2016}
given by
\begin{equation}
\tilde{\omega}_{s}(s)=\frac{\tilde{\omega}(0)}{\left[\frac{\tilde{\omega}(0)}{\tilde{\omega}(1)}-1\right]s+1},
\end{equation}
where $\tilde{\omega}^{\prime}(s)/\tilde{\omega}(s)^{2}=1/\tilde{\omega}(0)-1/\tilde{\omega}(1)$
is a constant. In this special protocol, the extra work by the sum
of Eqs (\ref{eq:mean extra work oscillator}) and (\ref{eq:oscillating extra work})
can approach zero at the specific control time $\tau=n\pi/\tilde{\varphi}(1),\,n=1,2,...$,
with the dynamical phase $\tilde{\varphi}(1)=[\ln(\tilde{\omega}(0))-\ln(\tilde{\omega}(1))]/[1/\tilde{\omega}(1)-1/\tilde{\omega}(0)]$.

\begin{figure}
\includegraphics[width=8cm]{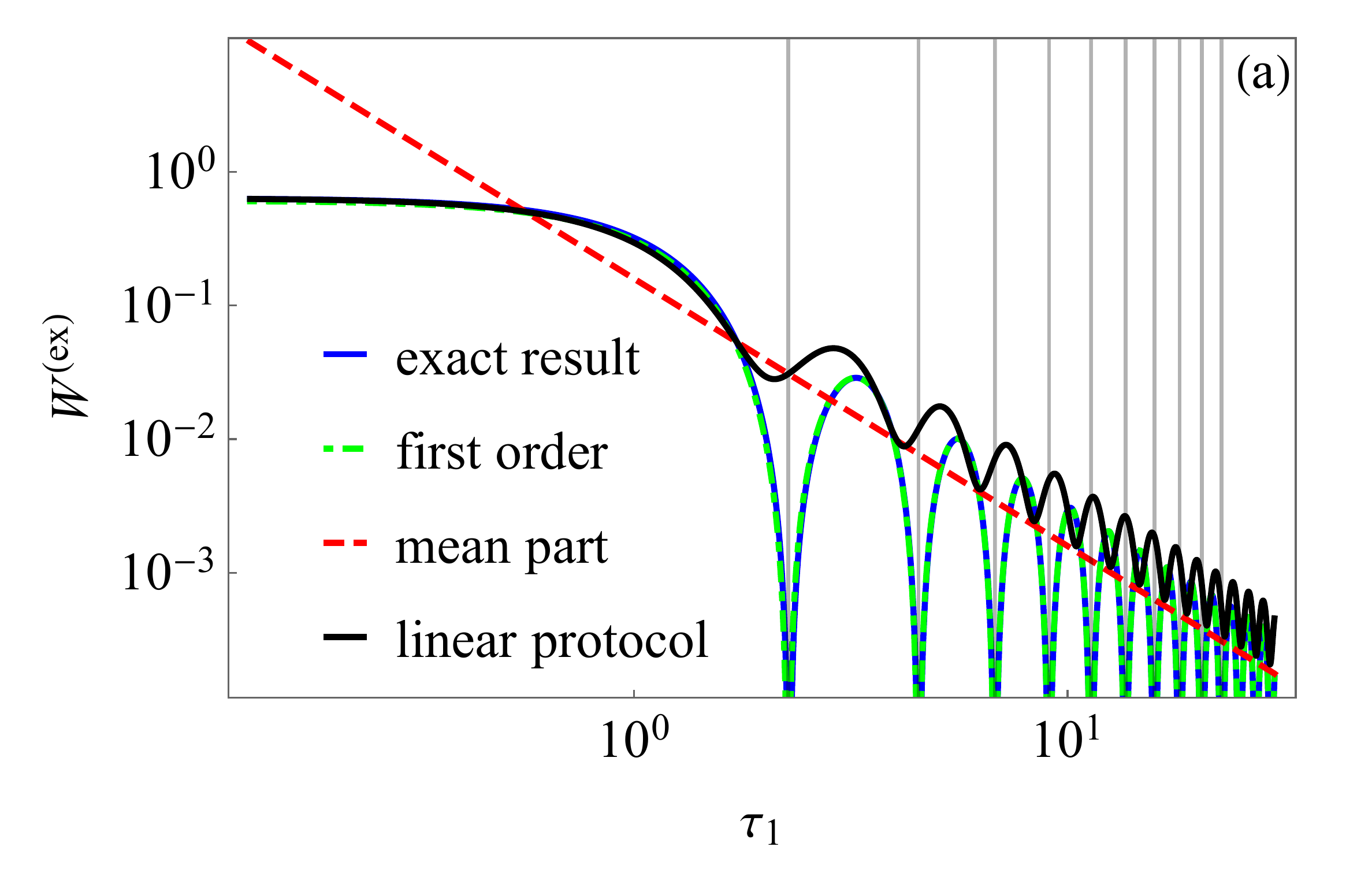}

\includegraphics[width=8cm]{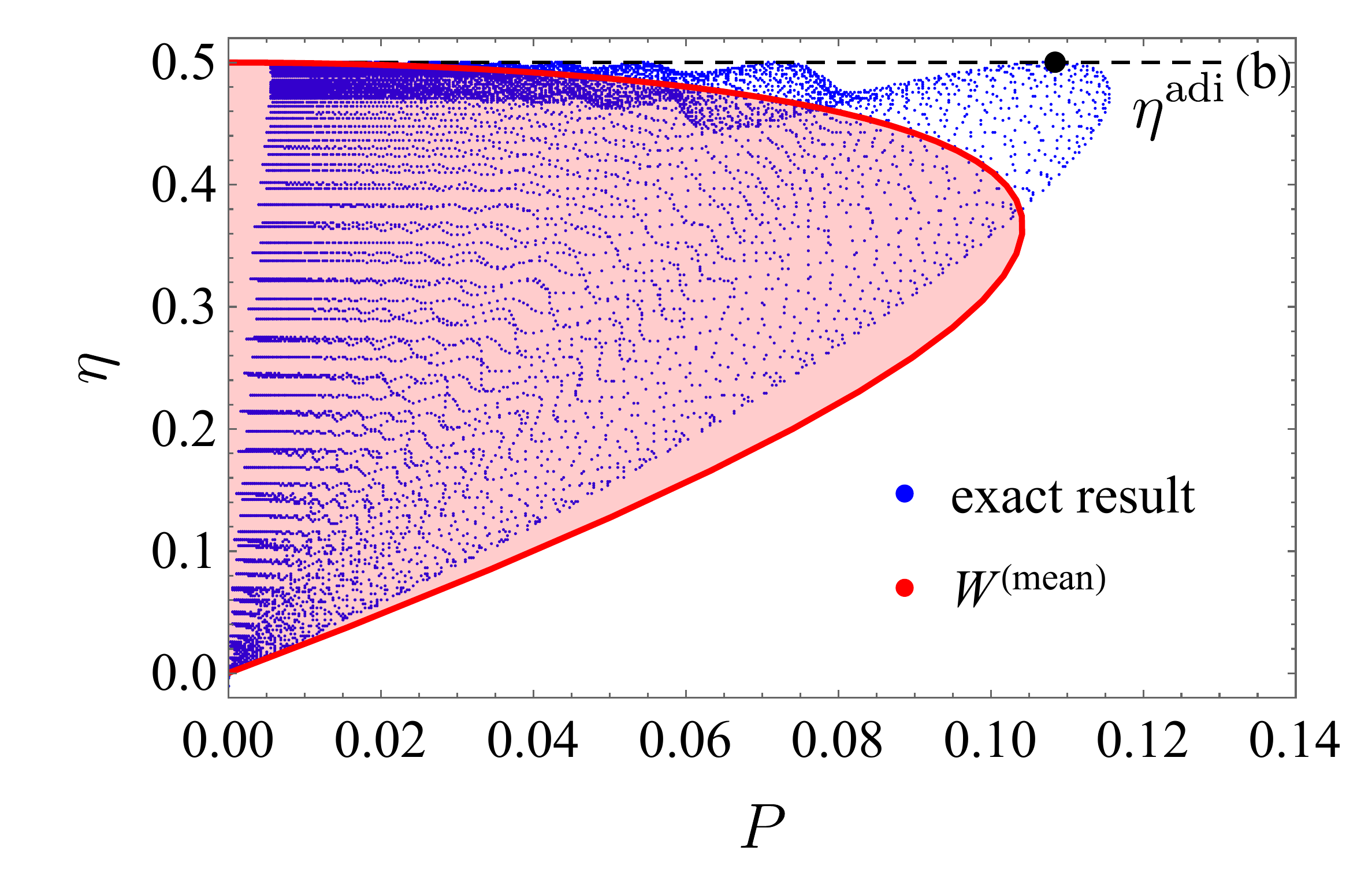}

\caption{\label{fig:specialprotocol-harmonic}(a) the extra work for the adiabatic
process $1\rightarrow2$ with the special protocol $\tilde{\omega}_{s}$
with different control time $\tau_{1}$. The vertical gray-dashed
line shows the extra work approaches zero at the specific control
time. (b) the reachable power and efficiency for the quantum harmonic
Otto engine. All the parameters are chosen the same in Fig. \ref{fig:comparing-harmonic}.}
\end{figure}

Figure \ref{fig:specialprotocol-harmonic} shows the results for the
special protocol, with the same parameters chosen in Fig. \ref{fig:comparing-harmonic}.
Figure \ref{fig:specialprotocol-harmonic}(a) presents the first-order
result (green-dashdotted curve) and the exact numerical result (blue-solid
curve) of the extra work, with the exact result of the linear protocol
shown as the black-solid curve for comparing. Figure \ref{fig:specialprotocol-harmonic}(a)
clearly shows that the extra work is smaller compared to that of the
linear protocol for most control time $\tau$, and can approach zero
at the specific control time.

Figure \ref{fig:specialprotocol-harmonic}(b) shows the constraint
between the efficiency and the power for the special protocol. When
the control time of the two adiabatic processes is chosen as the specific
control time $\tau=n\pi/\tilde{\varphi}(1)=n\pi/(2\ln2),\,n=1,2,...$,
the efficiency approaches to the one of the quasi-static Otto cycle
(the horizontal black-dashed line). For the specific control time
$\tau_{1}=\tau_{3}=\pi/(2\ln2)$, the heat engine gains large power
with the quasi-static efficiency, located as the black point. Compared
to the linear protocol in Fig. \ref{fig:comparing-harmonic}(b), the
quantum Otto engine with the special protocol attains larger maximum
power and the higher efficiency.

\section{Conclusion \label{sec:Conclusion}}

In this paper, we study the effect of the oscillating extra work on
both efficiency and output power for the quantum system with simple
energy-level structure, e. g. the two-level system and the quantum
harmonic oscillator. We conclude that the oscillating property of
the extra work can be utilized to obtain higher maximum power and
higher efficiency at the maximum power for the finite-time quantum
Otto engine by elaborately controlling the finite-time adiabatic processes.

We design special control schemes for the finite-time adiabatic process,
where the extra work approaches zero at the specific control time.
By adopting the special protocol in the finite-time Otto engine, the
engines can be optimized to approach the quasi-static efficiency $\eta^{\mathrm{adi}}$
with non-zero output power in finite-time Otto cycle.
\begin{acknowledgments}
HD would like to thantk D.Z. Xu for helpful discussion. This work
is supported by the NSFC (Grants No. 11534002 and No. 11875049), the
NSAF (Grant No. U1730449 and No. U1530401), and the National Basic
Research Program of China (Grants No. 2016YFA0301201 and No. 2014CB921403).
H.D. also thanks The Recruitment Program of Global Youth Experts of
China.
\end{acknowledgments}

Note: Upon finishing the current paper, we notice the implement of
quantum Otto cycle in experiment \citep{Peterson2018,Assis2019},
where the experimental result \citep{Peterson2018} clearly shows
the oscillation in power and efficiency with the increasing control
time. An analytical result of efficiency and power can be obtained
at long control time with our general formalism to match the numerical
and experimental results in Ref. \citep{Peterson2018}.

\appendix

\section{The Two-level system\label{sec:two-level system appendix}}

In this Appendix, we give the derivation of the asymptotic amplitude
to the first-order adiabatic approximation by Eq. (\ref{eq:cge=00005B1=00005D-1-1}).
Representing the amplitude $b_{nl}(s)=c_{nl}(\tau s)$ with the rescaled
time parameter $s$, Eqs. (\ref{eq:cng}) and (\ref{eq:cne}) are
rewritten as 
\begin{align}
\frac{\mathrm{d}}{\mathrm{d}s}b_{ng} & =e^{-2i\tau\tilde{\phi}(s)}\frac{\sin\left|\theta\right|}{2\tilde{\Lambda}(s){}^{2}}\frac{\mathrm{d}\tilde{\lambda}}{\mathrm{d}s}b_{ne}\\
\frac{\mathrm{d}}{\mathrm{d}s}b_{ne} & =-e^{2i\tau\tilde{\phi}(s)}\frac{\sin\left|\theta\right|}{2\tilde{\Lambda}(s){}^{2}}\frac{\mathrm{d}\tilde{\lambda}}{\mathrm{d}s}b_{ng}.
\end{align}
According to Ref. \citep{Chen2019}, the solution to the first order
of adiabatic approximation is carried out as

\begin{align}
b_{ge}^{[1]}(s) & =\frac{i\sin\left|\theta\right|}{4\epsilon\tau}\left[\frac{\tilde{\lambda}^{\prime}(s)e^{2i\tau\tilde{\phi}(s)}}{\tilde{\Lambda}(s){}^{3}}-\frac{\tilde{\lambda}^{\prime}(0)}{\tilde{\Lambda}(0){}^{3}}\right],\label{eq:cge=00005B1=00005D-1}\\
b_{eg}^{[1]}(s) & =\frac{i\sin\left|\theta\right|}{4\epsilon\tau}\left[\frac{\tilde{\lambda}^{\prime}(s)e^{-2i\tau\tilde{\phi}(s)}}{\tilde{\Lambda}(s){}^{3}}-\frac{\tilde{\lambda}^{\prime}(0)}{\tilde{\Lambda}(0){}^{3}}\right].\label{eq:ceg=00005B1=00005D-1}
\end{align}
The amplitude at the end of the adiabatic process by Eq. (\ref{eq:cge=00005B1=00005D-1-1})
follows immediately $c_{ge}^{[1]}(\tau)=b_{ge}^{[1]}(1)$.

\begin{figure}[h]
\includegraphics[width=7cm]{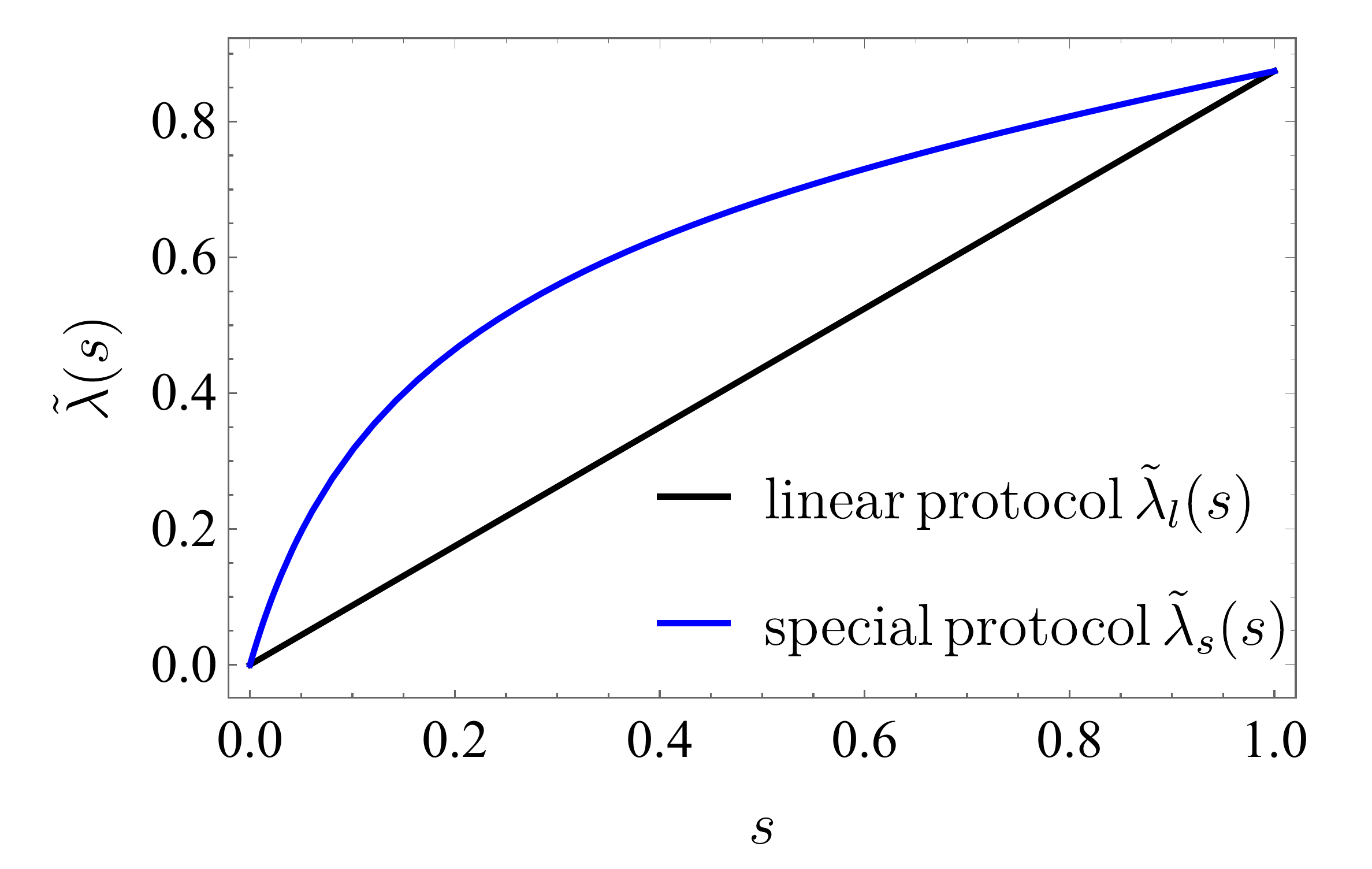}\caption{Two protocols for the two-level system, the lower black line for the
linear protocol $\tilde{\lambda}_{l}$ and the upper blue curve for
the special protocol $\tilde{\lambda}_{s}$. The parameters are chosen
as $\tilde{\lambda}(0)=0.1,\,\tilde{\lambda}(1)=0.8,\,\theta=0.4$.\label{fig:protocol}}
\end{figure}

Next, we give the explicit result for the special protocol $\tilde{\lambda}_{s}(s)$.
To allow the extra work approach zero at the specific control time,
we design a special protocol by setting $\tilde{\lambda^{\prime}}(s)/\tilde{\Lambda}(s){}^{3}=C$
as a constant at any moment during the adiabatic process. The constant
$C$ is determined by the initial $\tilde{\lambda}(0)$ and final
value $\tilde{\lambda}(1)$. Together with the initial and final condition,
we obtain the implicit function by Eq. (\ref{eq:specialprotocol}).
In Fig. \ref{fig:protocol}, we compare the special protocol $\tilde{\lambda}_{s}$
with the linear protocol $\tilde{\lambda}_{l}$, with the chosen parameters
$\theta=0.4,\,\epsilon=1,\,\tilde{\lambda}(0)=0.1,\,\tilde{\lambda}(1)=0.8$.
\begin{widetext}
For the the special protocol, we obtain the dynamical phase at the
end of the process as 
\begin{equation}
\tilde{\phi}(1)=\frac{\epsilon\sin(\theta)\left[\arctan\left(\frac{(1+\tilde{\lambda}(1))\tan\frac{\theta}{2}}{1-\tilde{\lambda}(1)}\right)-\arctan\left(\frac{(1+\tilde{\lambda}(0))\tan\frac{\theta}{2}}{1-\tilde{\lambda}(0)}\right)\right]}{\frac{\tilde{\lambda}(1)-\cos\theta}{\sqrt{\tilde{\lambda}(1)^{2}-2\tilde{\lambda}(1)\cos\theta+1}}-\frac{\tilde{\lambda}(0)-\cos\theta}{\sqrt{\text{\ensuremath{\tilde{\lambda}}(0)}^{2}-2\tilde{\lambda}(0)\cos\theta+1}}}.\label{eq:phitheta}
\end{equation}
The extra work approaches zero at the special control time $\tau=n\pi/\tilde{\phi}\left(1\right),\,n=1,2...$
\end{widetext}

\section{Time-dependent Harmonic Oscillator\label{sec:appendix Time-dependent-Harmonic-Oscillat}}
\begin{widetext}
In this appendix, we give the results of the time-dependent harmonic
oscillator, including the first-order adiabatic result and the method
for numerical calculation.

Following the method in Ref \citep{Chen2019}, the differential equation
of the amplitude $c_{nm}(t)$ follows from the Schrodinger equation
as

\begin{equation}
\frac{d}{dt}c_{nl}(t)+c_{nl}(t)\left\langle l\left|\dot{l}\right.\right\rangle +\sum_{m\ne l}c_{nm}(t)e^{-i(m-l)\varphi(t)}\left\langle l\left|\dot{m}\right.\right\rangle =0,
\end{equation}
where $\left|l\right\rangle =\left|l(t)\right\rangle $ is the instantaneous
eigenstate of the time-dependent harmonic oscillator. We rewrite the
equation with the rescaled time parameter $s$ as 
\begin{equation}
\frac{d}{ds}b_{nl}(s)+b_{nl}(s)\left\langle \tilde{l}\left|\frac{\partial}{\partial s}\right|\tilde{l}\right\rangle +\sum_{m\ne l}b_{nm}(s)e^{-i\tau(m-l)\tilde{\varphi}(s)}\left\langle \tilde{l}\left|\frac{\partial}{\partial s}\right|\tilde{m}\right\rangle =0,
\end{equation}
with $\left|\tilde{l}\right\rangle =\left|l(s\tau)\right\rangle $.
With the property of Hermite polynomial $H_{n}(\xi),$ we obtain the
derivative of the instantaneous eigenstate by Eq. (\ref{eq:eigenstate})
as
\begin{equation}
\left\langle \tilde{m}\right|\frac{\partial}{\partial s}\left|\tilde{n}\right\rangle =\frac{\tilde{\omega}^{\prime}(s)}{4\tilde{\omega}(s)}\left(-\sqrt{(n+1)(n+2)}\delta_{m,n+2}+\sqrt{n(n-1)}\delta_{m,n-2}\right).
\end{equation}
The terms $\left\langle \tilde{m}\right|\frac{\partial}{\partial s}\left|\tilde{n}\right\rangle $
with $m\ne n\pm2$ are all zero.

According to Ref. \citep{Chen2019}, we obtain the solution to the
first order of adiabatic approximation as

\begin{align}
b_{n,n+2}^{[1]}(s) & =-i\frac{\sqrt{(n+1)(n+2)}}{8\tau}\left(\frac{\tilde{\omega}^{\prime}(s)}{\tilde{\omega}(s)^{2}}e^{2i\tau\tilde{\varphi}\left(s\right)}-\frac{\tilde{\omega}^{\prime}(0)}{\tilde{\omega}(0)^{2}}\right),
\end{align}
and

\begin{equation}
b_{n,n-2}^{[1]}(s)=-i\frac{\sqrt{n(n-1)}}{8\tau}\left(\frac{\tilde{\omega}^{\prime}(s)}{\tilde{\omega}(s)^{2}}e^{-2i\tau\tilde{\varphi}\left(s\right)}-\frac{\tilde{\omega}^{\prime}(0)}{\tilde{\omega}(0)^{2}}\right).
\end{equation}
The diagonal term to the first order of adiabatic approximation is
$b_{n,n}^{[1]}(s)=1$ \citep{Chen2019}. The terms $b_{n,m}^{[1]}(s)=0$$,\,m\ne n,\,n\pm2$
are all zero since $\left\langle \widetilde{m}\right|\frac{\partial}{\partial s}\left|\tilde{n}\right\rangle =0$.
The amplitude at the end of the adiabatic process by Eqs. (\ref{eq:cnn+229},\ref{eq:cnn-230})
follows as $c_{n,n\pm2}^{[1]}(\tau)=b_{n,n\pm2}^{[1]}(1)$.
\end{widetext}

In Ref. \citep{Chen2010PhysRevLett104_63002}, the exact result of
the internal energy during the finite-time adiabatic process is described
by the non-adiabatic factor $\mathscr{N}(t)$ as

\begin{equation}
\left\langle H(\omega(t))\right\rangle =\frac{\omega(t)}{2}\mathscr{N}(t)\coth\left[\frac{\beta\omega(0)}{2}\right].\label{eq:internal energy}
\end{equation}
The non-adiabatic factor $\mathscr{N}(t)$ is determined by a scalar
$c(t)$ as 
\begin{equation}
\mathscr{N}(t)=\frac{\left[\dot{c}(t)\right]^{2}+\left[\omega(t)\right]^{2}\left[c(t)\right]^{2}+\frac{\left[\omega(0)\right]^{2}}{\left[c(t)\right]^{2}}}{2\omega(t)\omega(0)},\label{eq:qfactor}
\end{equation}
where $c(t)$ satisfies the differential equation

\begin{equation}
\text{\ensuremath{\ddot{c}(t)+\omega(t)^{2}c(t)=\frac{\omega(0)^{2}}{c(t)^{3}}}},\label{eq:c(t)harmonic}
\end{equation}
with the initial condition $c(0)=1,\,c^{\prime}(0)=0$.

With Eq. (\ref{eq:internal energy}), the work during the finite-time
adiabatic process is rewritten with $\mathscr{N}(t)$ as $W(\tau)=\frac{\omega(\tau)}{2}\mathscr{N}(\tau)\coth\left(\frac{\beta\omega(0)}{2}\right)-\frac{\omega(0)}{2}\coth\left(\frac{\beta\omega(0)}{2}\right)$.
Correspondingly, the quasi-static work is
\begin{equation}
W^{\mathrm{adi}}=\frac{\omega(\tau)-\omega(0)}{2}\coth\left(\frac{\beta\omega(0)}{2}\right),
\end{equation}
and the extra work is
\begin{equation}
W^{(\mathrm{ex})}(\tau)=\frac{\omega(\tau)}{2}\left[\mathscr{N}(\tau)-1\right]\coth\left(\frac{\beta\omega(0)}{2}\right).\label{eq:extra work Q}
\end{equation}
It is verified $\mathscr{N}(\tau)\geq1$, which approach $1$ for
infinite control time $\tau\rightarrow\infty$. The difference $\mathscr{N}(\tau)-1$
describes the non-adiabatic effect, and does not depend on the initial
inverse temperature $\beta$.
\begin{widetext}
Substituting Eq. (\ref{eq:extra work Q}) into Eqs. (\ref{eq:power})
and (\ref{eq:efficiency}), we rewrite the output power

\begin{equation}
P^{\mathrm{h}}=\frac{\left[\omega_{0}-\omega_{1}\mathscr{N}_{1}(\tau_{1})\right]\coth\left(\frac{\beta_{\mathrm{h}}\omega_{0}}{2}\right)+\left[\omega_{1}-\omega_{0}\mathscr{N}_{3}(\tau_{3})\right]\coth\left(\frac{\beta_{\mathrm{c}}\omega_{1}}{2}\right)}{2(\tau_{1}+\tau_{3})},\label{eq:powerharmonic}
\end{equation}
and the corresponding efficiency
\begin{equation}
\eta^{\mathrm{h}}=1-\frac{\omega_{1}\left[\mathscr{N}_{1}(\tau_{1})\coth\left(\frac{\beta_{\mathrm{h}}\omega_{0}}{2}\right)-\coth\left(\frac{\beta_{\mathrm{c}}\omega_{1}}{2}\right)\right]}{\omega_{0}\left[\coth\left(\frac{\beta_{\mathrm{h}}\omega_{0}}{2}\right)-\mathscr{N}_{3}(\tau_{3})\coth\left(\frac{\beta_{\mathrm{c}}\omega_{1}}{2}\right)\right]},\label{eq:efficiencyharmonic}
\end{equation}
where $\mathscr{N}_{1}(\tau_{1})$ and $\mathscr{N}_{3}(\tau_{3})$
denote the non-adiabatic factors for the two finite-time adiabatic
processes. In the numerical calculation, we first choose different
control time to obtain the exact result of the non-adiabatic factor
$\mathscr{N}_{i}(\tau_{i}),\,i=1,3$ for the two finite-time adiabatic
processes by solving Eq. (\ref{eq:c(t)harmonic}) numerically. Then,
we use Eqs. (\ref{eq:powerharmonic}) and (\ref{eq:efficiencyharmonic})
to calculate the exact power and efficiency, respectively.
\end{widetext}

\bibliographystyle{apsrev4-1}
\bibliography{main}

\end{document}